\documentclass[sigconf]{acmart}

\AtBeginDocument{%
  \providecommand\BibTeX{{%
    \normalfont B\kern-0.5em{\scshape i\kern-0.25em b}\kern-0.8em\TeX}}}

\setcopyright{acmcopyright}
\copyrightyear{2023}
\acmYear{2023}
\acmDOI{XXXXXXX.XXXXXXX}

\acmConference[SUI '23]{ACM Symposium on Spatial User Interaction}{October 13--15, 2023}{Sydney, Australia}
\usepackage{subcaption}
\usepackage{balance}

\begin{document}


\title{Immersive Virtual Reality and Robotics for Upper Extremity Rehabilitation}


\author{Vuthea Chheang, Rakshith Lokesh, Amit Chaudhari, Qile Wang, Lauren Baron, Behdokht Kiafar, Sagar Doshi, Erik Thostenson, Joshua Cashaback, Roghayeh Leila Barmaki 
}
\affiliation{
    \institution{
    \vspace{0.2cm}
    Department of Computer and Information Sciences, University of Delaware} \country{Newark, DE, United States}
}

\renewcommand{\shortauthors}{Chheang et al.}

\begin{abstract}
Stroke patients often experience upper limb impairments that restrict their mobility and daily activities. Physical therapy (PT) is the most effective method to improve impairments, but low patient adherence and participation in PT exercises pose significant challenges. To overcome these barriers, a combination of virtual reality (VR) and robotics in PT is promising. However, few systems effectively integrate VR with robotics, especially for upper limb rehabilitation. 
This work introduces a new virtual rehabilitation solution that combines VR with robotics and a wearable sensor to analyze elbow joint movements. The framework also enhances the capabilities of a traditional robotic device (KinArm) used for motor dysfunction assessment and rehabilitation. A pilot user study (n = 16) was conducted to evaluate the effectiveness and usability of the proposed VR framework. We used a two-way repeated measures experimental design where participants performed two tasks \textit{(Circle and Diamond)} with two conditions \textit{(VR and VR KinArm)}. 
We observed no significant differences in the main effect of conditions for task completion time. However, there were significant differences in both the normalized number of mistakes and recorded elbow joint angles (captured as resistance change values from the wearable sleeve sensor) between the \textit{Circle} and \textit{Diamond} tasks. Additionally, we report the \textit{system usability, task load}, and \textit{presence} in the proposed VR framework. This system demonstrates the potential advantages of an immersive, multi-sensory approach and provides future avenues for research in developing more cost-effective, tailored, and personalized upper limb solutions for home therapy applications. 
\end{abstract}

\ccsdesc[500]{Human-centered computing~Usability testing}
\ccsdesc[500]{Human-centered computing~Virtual reality}
\ccsdesc[500]{Software and its engineering~Interactive games}
\keywords{Virtual reality, upper extremity, rehabilitation, wearable sensors, end-point robots, human-computer interaction}

\begin{teaserfigure}
  \includegraphics[width=\textwidth]{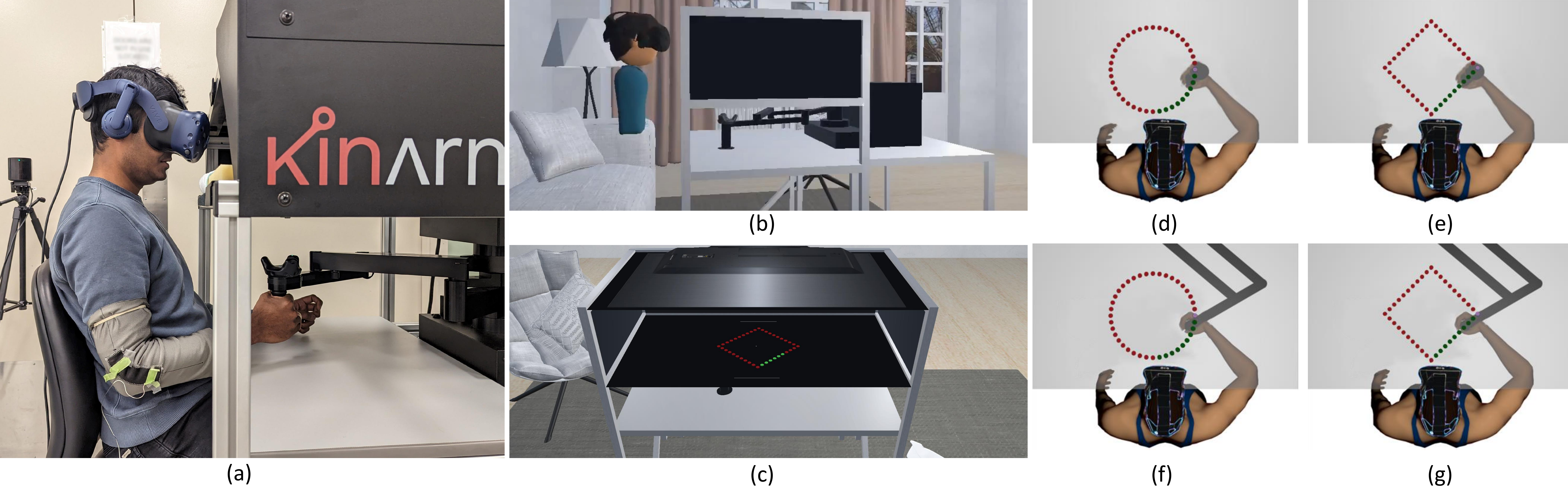}
  \caption{Overview of the therapeutic system for upper extremity rehabilitation using immersive VR and end-point robotics (KinArm): (a) technical setup for study conditions, (b) the VR environment with virtual avatar and virtual robotics with inverse kinematics, (c) the first-person view in VR, and conceptual views of the conditions and tasks: VR condition with the (d) circle and (e) diamond tasks, and VR KinArm robotic condition with the (f) circle and (g) diamond tasks.}
  \Description{Overview of the therapeutic system  for upper extremity rehabilitation using immersive VR and end-point robotics.}
  \label{fig:teaser}
\end{teaserfigure}


\maketitle


\section{Introduction}

Stroke is one of the major causes of disability, and more than 75\% of stroke survivors have significant motor impairments \citep{hatem2016rehabilitation}. The most common case after stroke is hemiparesis of the upper limb, including muscle weakness, joint laxity, and impaired motor control \citep{mozaffarian2015heart}. 
The upper extremity impairment induces disabilities for everyday tasks and activities, e.g., reaching, holding, and picking objects \citep{bleyenheuft2014precision, hatem2016rehabilitation}. 
Moreover, previous research has shown that regaining lost functions in the upper extremities is often more complicated than those in the lower extremities \citep{iruthayarajah2016upper, welmer2008limited}.
Physical therapy and therapeutic exercises are highly recommended for rehabilitation and improving quality of life \citep{ingram2021quantifying, seymour1993aftermath}. Physical therapy is usually performed under the supervision of a trained therapist and in a clinical setting. The patients are also expected to perform regular exercises at home. However, studies have shown that the patients' adherence levels to the prescribed regimens are often low because of the inability to track progress, lack of motivation, slow recovery progress, and limited understanding of the benefits of treatment \citep{emmelkamp2021virtual, wiederhold2019virtual}. 

The increasing progress in technology and computational power has led to a rise in the use of virtual reality (VR) for physical therapy, which has the potential to enhance patient adherence.
Patients suffering from neuro-musculoskeletal injuries, including stroke, need to perform reaching and stretching movements to recover mobility in their upper limbs. 
The use of VR in upper limb therapy offers several benefits, such as increased patient engagement through game-like characteristics, portable setup for home therapy, reduced outside pressures and distractions, the ability to attenuate pain, and the ease of measuring and evaluating performance \citep{juan2022immersive, baron2021enjoyable}.

Compared to traditional physical therapy, robotic rehabilitation also promotes remarkable improvement in motor and physiological skills \citep{wu2021robot, frisoli2007arm}.  
However, investigation on integrating immersive VR with end-point robots, especially for upper extremity rehabilitation, is still underrepresented \citep{tarnita2022analysis, mubin2019exoskeletons}.   
Additionally, the current upper extremity VR rehabilitation tools mainly focus on the analysis of hand movements \citep{8826947}.
Thus, an assessment of additional factors pertaining to upper limb exercises during VR therapy is needed. 
For example, testing how VR-based settings influence user (patient) performance and how rehabilitation tasks can be customized or used to improve user engagement.


In this work, we present a framework for upper extremity rehabilitation using immersive VR and end-point robotics (see~\autoref{fig:teaser}). 
We have developed a VR environment with two reaching tasks and replicated the end-point robot in the virtual setting.
A knit fabric-based nanocomposite sensor was used to assess elbow joint movement in addition to the hand movement data from VR/KinArm for a holistic assessment of the upper extremities. 
This sensor was sewn into a one-size-fits-all sleeve for the user to wear on their dominant arm.

We conducted a pilot user study (n = 16) to assess the effectiveness of our framework from the perspective of user performance, usability, task load, and sense of presence. The results provide insights regarding the potential benefits of our VR-based system for upper extremity rehabilitation. They also show future research directions for virtual therapy and multi-sensor movement sensing for more precise and data-driven therapeutic assessment.
Our contributions are the following:
\begin{itemize}

    \item Development of a therapeutic VR framework that combines VR with KinArm robotics and a wearable sleeve sensor to capture and measure multimodal sensing information for rehabilitation assessment.
    
    \item Results of a pilot user study ($n = 16$) providing insights on usability, task load, presence, and user performance as a preliminary system assessment before extensive clinical evaluation on the target group. 
    
    \item Exploratory analysis of participants' qualitative feedback to determine advantages, limitations, and potential improvements and research directions.

\end{itemize}


\section{Related Work}


In this section, we describe related work regarding the use of VR, wearable sensors, and end-point robotics for upper extremity rehabilitation.

\subsection{Virtual Therapy and Wearable Sensors} 

VR has been shown to improve the quality, adherence, and mental health of patients \citep{hoeg2023buddy, baron2021enjoyable}. 
A recent review reported the feasibility of using VR in  treating 
obsessive-compulsive disorder, psychosis, autism spectrum disorder, attention deficit hyperactivity disorder, post-traumatic stress disorder, addiction, and eating disorders \citep{emmelkamp2021virtual}.
Several other studies have supported virtual therapy for social anxiety disorders, specifically with virtual exposure therapy \citep{emmelkamp2020virtual, morina2015can, kampmann2016exposure}.
Cognitive behavior therapy has also been shown to successfully treat mental disorders by integrating clinical VR technology \citep{lindner2021better}. 

VR-based physical therapy gives patients support and flexibility to exercise in a home environment with fewer visits to clinics \citep{hoeg2020co}. 
\citet{cui2019wearable} proposed a self-guided healthcare system, with a wireless inertial sensor combined with VR interactive technology. This system provides a frozen shoulder joint mobility self-measurement system that can measure shoulder joint movements at-home without the intervention of a specialist. 
The feedback provided by the inertial sensor is critical in wireless remote data collection for the patient's shoulder joint angle and position. 
The results from patient interviews showed that the patients with minimal technical knowledge were confident and willing to operate the proposed system to enhance their understanding of their rehabilitation progress. 
\citet{herrera2023rehab} developed a clinical Rehab-Immersive (RI) framework to support the rehabilitation of patients with spinal cord injuries addressing upper limb motor impairments. They found that RI has the potential to enhance the rehabilitation experience for both therapists and patients.

For VR-based physical therapy systems, wearable sensors are critical for monitoring physical activity and providing feedback remotely for the physical activities/exercise performed in a home setting. 
\citet{rawashdeh2022highly} used an inertial measurement unit on the body to capture the patient's form during exercise, and the kinematics data processing approach is used to define custom exercises as well as real-time corrective feedback parameters.  
Integration of wearable sensors with a virtual physical therapy system gives an edge over traditional rehabilitation methods by offering real-time feedback to the patient, making it more engaging and enabling 
remote rehabilitation in a home setting. 
\citet{8826947} used IoT wearable devices developed 
by embedding smart sensors in headbands and two gloves and proposed a VR-based solution for the physical rehabilitation of upper limbs. 
VR serious games are used for finger rehabilitation, and a Web API was used to store and process sensor data to evaluate physical capabilities and limitations during exercise. This system can engage the patients in their training sessions and help therapists optimize the training plans.

\begin{figure}[!t]
    \centering
     \includegraphics[width=\columnwidth]{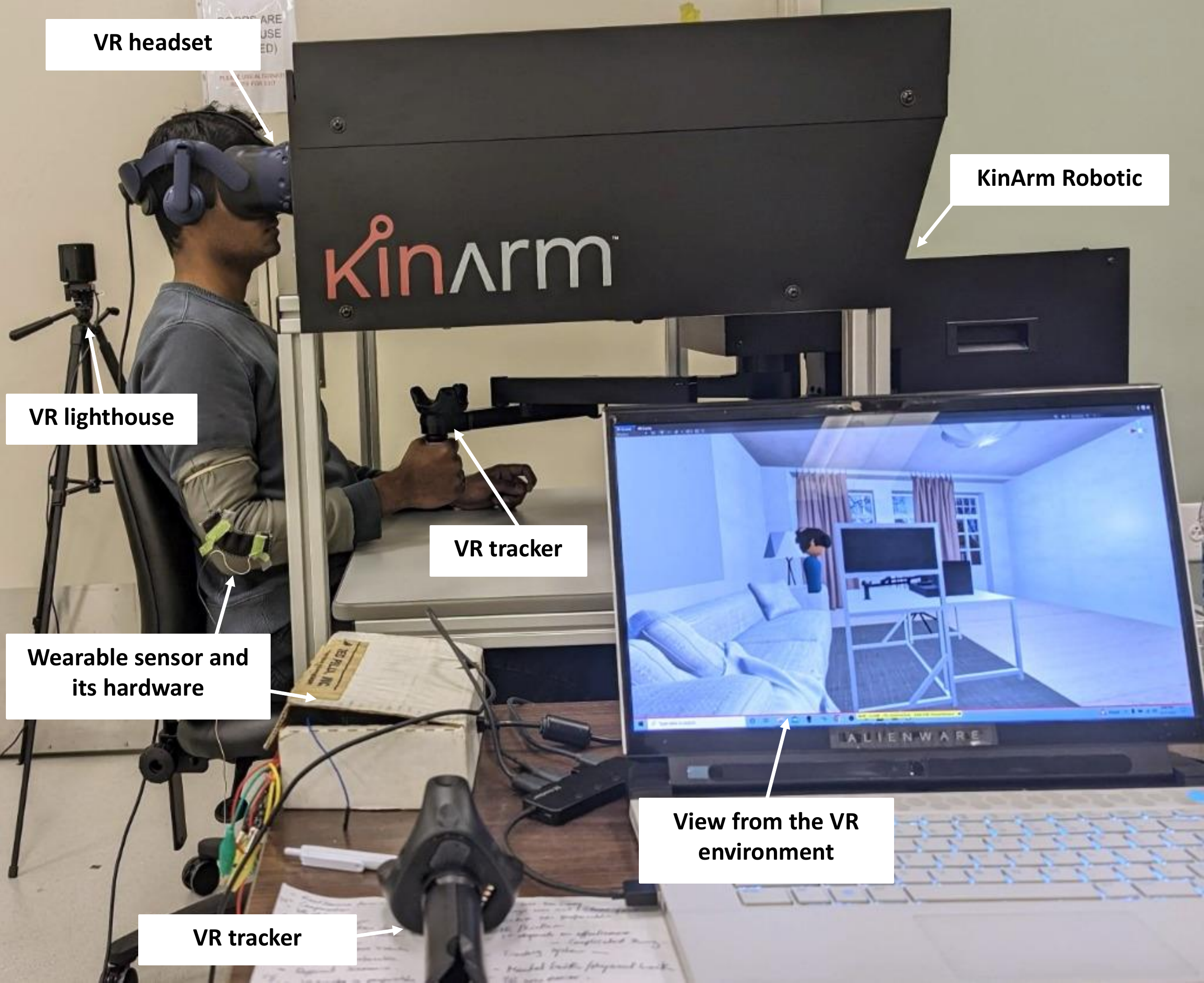}
    \caption{Technical setup of the VR upper rehabilitation and robotics.}
    \label{fig:technicalsetup}
\end{figure}

\subsection{End-Point Robotics for Upper Extremity Rehabilitation}

Robot-mediated neurorehabilitation has attracted significant attention in the last decade due to advancements in the field of robotics and neuroscience.
\citet{walker2022virtual} provided a review of systems using robots with VR and augmented reality (AR) to enhance efficiency and safety in healthcare and also being utilized in therapeutic and assistive robots.
The robots are primarily applied to exercise the impaired limbs to restore neurological and sensorimotor function. 
State-of-the-art rehabilitation robots are capable of measuring limb movements and reliably controlling movements at the joint level \citep{gassert2018rehabilitation}.
Computational motor learning principles and advanced training algorithms have also been integrated to provide customized support to each patient \citep{huang2009robotic, patton2004robot, marchal2009review}. 

In comparison to conventional physical therapy, rehabilitation robots can be used to administer consistent and high-intensity training for a longer duration of time \citep{huang2009robotic, lo2012exoskeleton}. 
Several robotic rehabilitation studies have reported significant improvements in motor control and physiological measures in individuals affected by stroke \citep{kwakkel2008effects, wu2021robot}. 
\citet{hussain2019end} developed a rehabilitation system that use adaptive robotics with VR simulations specifically designed to assist children with upper extremity hemiplegia.
Creating fully-immersive VR-based environments that closely resemble real-life scenarios could be instrumental \citep{ona2019robotics}.
In addition, full functional recovery is possible only when assessment and rehabilitation can be performed outside of the lab/clinic for various activities of daily living \citep{ gassert2018rehabilitation}.
A viable solution for providing functional assessments and training at home has been the use of smart wearable devices.

\citet{wonsick2020systematic} presented a systematic review indicating the rapid growth and synergy between VR interfaces and the field of robotics. 
Particularly, individuals suffering from motor impairments caused by stroke or Parkinson's disease can greatly benefit from such technology. 
A desktop-based system was developed to rehabilitate hand function in stroke patients by using two input devices: a cyber-glove and a \textit{Rutgers Master II-ND} force feedback glove which enables patients to interact with a virtual environment \citep{jack2001virtual}. Their clinical trials have shown improvement in most of the hand parameters and the patients had positive subjective evaluations.
For patients with Parkinson's disease, a meta-analysis found that VR-based therapy can enhance balance and gait \citep{wang2019effect}. 

\citet{mohammadi2018real} presented a framework for using VR and robots to enable juggling games for patients in motor rehabilitation to release the burden on therapists and improve the effectiveness of physiotherapy.
Their system combines real-time motion generation based on targeting human motion, real-time robot control, and VR for therapeutic juggling. The study highlights that there are still crucial questions to be answered and future studies are needed to advance our knowledge about human-robot interaction. 
Another systematic review \citep{baur2018trends} evaluated the use of multiplayer interactions in rehabilitation therapy for patients with cognitive and/or motor impairments. 
The study found that multiplayer training for rehabilitation improved the overall experience and user performance by providing haptic support or haptic resistance.

\section{Materials and Methods}

The experiment was approved by the institutional review board. 
We describe the participants, apparatus, study procedure, and study design in the following sections. 

\subsection{Participants}

We conducted \emph{a priori} power analysis to determine the sample size for interaction effects with ANOVA analysis (repeated measures, within factors). A statistical G*Power analysis program was used with an effect size $\eta_{p}^{2}: 0.14$ for two groups, resulting in a total sample size of 16 \citep{faul2007g}.
Since it is an early prototype, the aim was to evaluate the feasibility, usability, and qualitative feedback with non-clinical subjects. 
Therefore, 16 participants were recruited for the study (8 males and 8 females). The participation was voluntary, and there was no monetary compensation.  
 They were between 22 and 37 years old (\textit{M}\,=\,27.38, \textit{SD}\,=\,4.13). 
Six of them had a background in computer science, three in business analytics, and others in education, electrical engineering, geography, biomedical engineering, mechanical engineering, environmental policy, and political science.
One was in the Bachelor's, six in the Master's, and nine in the doctoral program.
Among all participants, nine reported having video game experience (six reported playing several times a year and three reported playing it daily).
Seven participants had previous VR experience. They reported that they used it several times a year. All participants were right-handed.


\subsection{Apparatus}

A virtual environment for upper extremity rehabilitation was developed with the \textit{Unity} game engine (version 2021.3.10f1). We used \textit{VRTK} toolkit for basic interactions in VR, e.g., teleportation. The built-in \textit{OpenXR} package in Unity was used to make the VR environment compatible with several VR headsets. 
In this study, we used \textit{HTC Vive Pro Eye}, tracker, and its components for VR tracking space.
For the virtual scene, the living room and additional models from \textit{Sketchfab} were customized and integrated into the virtual environment. 
The user was represented as an avatar with their virtual hands in the VR environment. 
The idea was to provide a comfortable and motivated environment for patients, especially those with limited mobility.     

We used a KinArm end-point robot (\textit{BKIN Technologies Ltd.}, Canada) integrated with the VR setup (see~\autoref{fig:technicalsetup}). The end-point robot is a planar manipulandum that consists of a cylindrical handle. Participants could grasp the handle with their hand and move the handle freely in the horizontal plane (i.e. 2 dimensions, lateral and forward). 
A virtual model for the KinArm robot was designed to mimic the physical one in the virtual environment (see also~\autoref{fig:teaser}). 
The HTC Vive tracker was attached to the KinArm robotic handle to get its movement position and simulate the inverse kinematics for the VR KinArm setup. Thus, when the user moves the KinArm robotic handle, the movement of the virtual KinArm is simulated too.

To evaluate the joint movement at their elbow, a wearable fabric-based carbon nanotube sensor \citep{doshi2022ultrahigh} was used. The sensor was made of knit fabric, and it was integrated into a fabric sleeve. 
This sensor is piezo-resistive and resistance across the electrode changes with the stretch/strain in the sensor for elbow flexion/extension. We used this wearable sensor on the participant’s elbow for all the conditions (see also \autoref{fig:technicalsetup}). 
The resistance was recorded by an Arduino-based voltage divider circuit and data was streamed via a \textit{Microsoft Excel} streamer. The sensor provided changes in electrical resistance corresponding to elbow flexion/extension during the performance.

\subsection{Study Procedure}

\begin{figure*}[t]
    \centering
    \includegraphics[width=\textwidth]{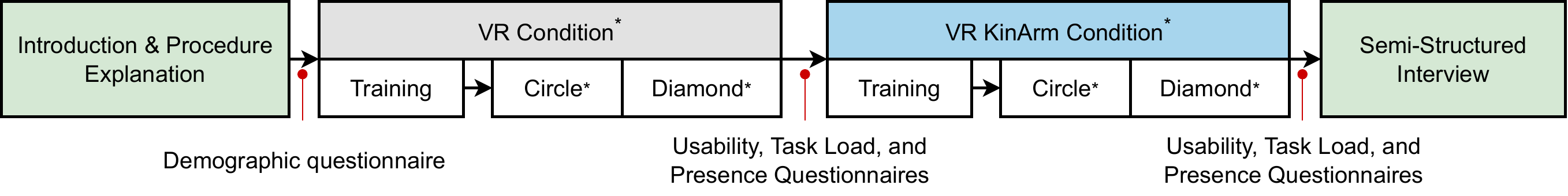}
    \caption{Overview of the study procedure. The order of the conditions (marked with $^*$) was counterbalanced.}
    \label{fig:procedure}
\end{figure*}

An overview of the study procedure is shown in \autoref{fig:procedure}.
The participants were welcomed with a brief overview of the project and study. They were asked to complete a demographic questionnaire. 
Then, they were informed about the detailed study procedure and were instructed in the use of the KinArm robot, VR headset, and sleeve sensor. 
The entire session for each participant was planned for one hour.

The procedure started with a training in a different virtual environment, and they were asked to perform a simple task. They were also asked to explore all the features and interaction possibilities. The tasks and the measurements were explained. If the participants stated that they understand the procedure, they were connected to the actual VR environment with the first condition, and their performance data was recorded.  
The order of the conditions was counterbalanced.

After completing the first condition, the participants were asked to take off the VR headset and answer mid-questionnaires, that included usability, presence, and task load index questionnaires. They were asked to rest for a few minutes before performing the second session.
After that, the procedure for the second condition was initiated, and the same training and experimental protocols were followed. 
Lastly, after the participants completed all the conditions, we conducted a semi-structured interview to collect feedback regarding their experience with the system.

\subsection{Study Design}

This study is a $2\times2$ with-in subjects design. Here we describe the design of our user study, including independent variables, dependent variables, subjective measures from questionnaires, and semi-structured interviews.
Our research questions are the following:

\begin{itemize}
    \item [\textbf{RQ1}] How do \textit{VR} and \textit{VR KinArm} robotics conditions influence the user performance for therapeutic tasks?
    
    \item [\textbf{RQ2}] How are subjective measures, including usability, task load, and presence, associated with both conditions?
    
    \item [\textbf{RQ3}] What are the applicability, perceived advantages, and limitations of each condition?

\end{itemize}

\subsubsection{Independent Variables}

The user study was planned as a within-subject design with a two-factor test. These two factors were defined as independent variables: \textit{Condition} and \textit{Task} (see \autoref{fig:teaser}). 
The \textit{Condition} refers to the therapeutic setup, and there were two conditions: \textit{VR and VR KinArm} (see also \autoref{fig:teaser}). 

\begin{itemize}
    \item \textbf{VR}: the participants were asked to use a VR tracker from HTC Vive to perform the tasks. They were asked to hold and move it on the table. Two lighthouses were used to track the position of the VR tracker and headset and to calibrate the participants in the virtual environment.      
    
    \item \textbf{VR KinArm}: this condition is an integration of the VR setup and the KinArm robot. The participants were asked to use the KinArm robot with its handle to perform the task. We placed the VR tracker on the handle of the KinArm to track its position and to simulate the virtual KinArm in the virtual environment. 
    
\end{itemize}

There were two \textit{Tasks} for each condition: \textit{Circle} and \textit{Diamond}. These \textit {Tasks} were designed to have a similar level of complexity and realism according to therapeutic tasks provided by the KinArm exoskeleton.
The participants were asked to follow a pink and blinking dot (see also \autoref{fig:teaser}).   
If the dot is selected, it turns green while the remaining dots are in red. To avoid bias due to the learning effect, we followed a balanced Latin square \citep{keedwell2015latin}. Therefore, the order of the tasks and the direction of drawing (clockwise and counterclockwise directions) were counterbalanced.  

\begin{itemize}
    \item \textbf{Circle}: the drawing content was an abstract outline as a circle shape with 40 drawing dots.  
    \item \textbf{Diamond}: the drawing content was an abstract outline as a diamond shape with 40 drawing dots. 
\end{itemize}

An abstract outline aims to provide a simple task but a better assessment of the wearable sleeve sensor for resistance changes from elbow stretches.
The diameter of each dot was 5\,cm, and it was identical for both tasks for consistency.
Both tasks encourage broad arm movements to mimic upper extremity exercises; the circle task encourages more continuous arm movement, while the diamond encourages more rigid movements given the right angles in the shape. 

\subsubsection{Dependent Variables}

Three measurements to objectively evaluate the drawing performance are task completion time, number of mistakes, and resistance change from the wearable sleeve sensor.

\begin{itemize}
    \item \textbf{Task Completion Time}: the completion time of the drawing task was calculated based on the starting time of the task until the participant hit all of the dots.

    \item \textbf{Normalized Number of Mistakes}: the number of mistakes relative to the task completion time (i.e., number of mistakes per second). 
    For example, if successive dots were not hit, the number of missed dots was divided by the task completion time. 

    \item \textbf{Resistance Change}: 
    the change in electrical resistance of the sleeve sensor was measured during elbow joint flexion and extension while performing the task.

\end{itemize}

\subsubsection{Questionnaires}

Besides the performance data, we also collected the questionnaire data as subjective measures. All questionnaires were designed using the Qualtrics survey platform. 

\begin{itemize}
    \item \textbf{Usability}: the system usability scale (SUS) questionnaire was used to assess usability \citep{Brooke1995SUS}. This questionnaire includes ten questions with a 5-point Likert-scale ranging from \textit{strongly disagree} to \textit{strongly agree}.
    The SUS score was then converted to a range between 0--100\% (0--50\%: not acceptable, 51--67\%: poor, 68\%: okay, 69--80\%: good, 81--100\%: excellent) \citep{bangor2009determining}. 

    \item \textbf{Task Load}: \textit{NASA TLX} questionnaire was used as an indicator to evaluate the subjective task load of the system interactions and conditions \citep{hart2006nasa}. The questionnaire includes
    mental demand, physical demand, temporal demand, performance, effort, and frustration.  
    
    \item \textbf{Presence}: to evaluate the sense of presence in the immersive environment, \textit{igroup presence questionnaire} (IPQ) was used \citep{schubert2001experience, schwind2019using}. The questionnaire contains 14 questions with 7-point Likert-scale ranging from \textit{strongly disagree} to \textit{strongly agree}. It is divided into four categories: general presence, spatial presence, involvement, and experienced realism.

\end{itemize}

\subsubsection{Semi-Structured Interviews}

After completion of all the conditions, we conducted the semi-structured interview to collect participant feedback.
The following questions were asked during the interview:

\begin{itemize}
    \setlength\itemsep{0em}
    \item What is your feedback 
    of the tasks and conditions?
    \item Do you have any questions or suggestions?
\end{itemize}

\subsubsection{Data Analysis}

\textit{RStudio} with R was used for data analysis.
A two-way analysis of variance (ANOVA) was conducted for dependent variables. 
For resistance change of wearable sleeve sensor, data for one participant was excluded from the analysis due to very high values of resistance, possibly due to error in connection/ measurement.
The performance and questionnaire data were further analyzed with pairwise t-tests and the Bonferroni adjustment method to determine the differences between the conditions. 
For participant feedback, the comments were collated into a database, and the redundancies were removed.

\section{Results}

In the following sections, we describe the results of statistical analysis for user performance, questionnaire results, and qualitative participant feedback.

\subsection{User Performance Results (RQ1)}

\begin{figure*}[!t]
    \centering    \includegraphics[width=\textwidth]{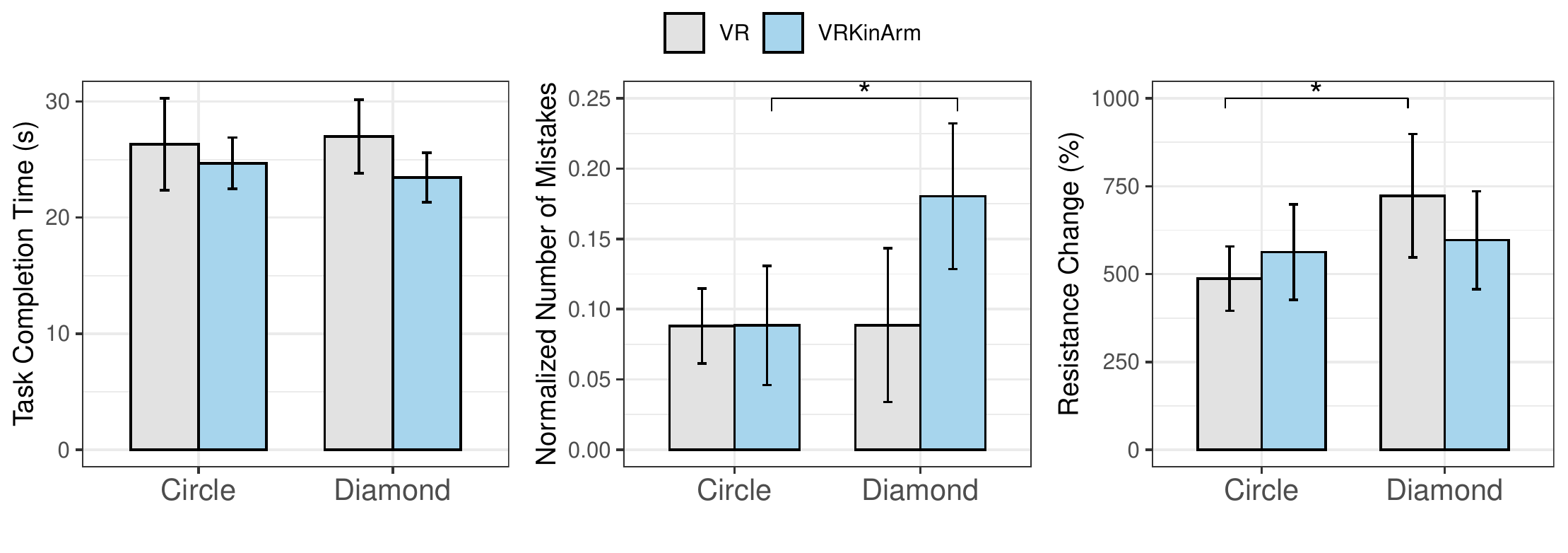}
    \caption{Results of objective measures resulted from the user performance (n=16). * denotes significance.}
    \label{fig:performanceResults}
\end{figure*}

The summary of descriptive results for objective measures of user performance is shown in \autoref{tab:descriptiveResults} and \autoref{fig:performanceResults}. 
The statistical results are listed in \autoref{tab:anovaresults}.

\subsubsection{Task Completion Time (TCT)}
We did not find any significant differences in the task completion time on their main effects for both factors: \textit{Condition (VR and VR KinArm)} 
and \textit{Task (Circle and Diamond)}. 
There was also no significant effect for the interaction effect. 
It could indicate that both conditions are comparable and interchangeable regarding the task completion time.
For descriptive results, the \textit{VR KinArm Condition} (\textit{M} = 24.10, \textit{SD} = 8.59)
was on average faster than the \textit{VR Condition} (\textit{M} = 26.60, \textit{SD} = 14.10). 
However, the results show a small difference between these conditions.

\begin{table*}[!t]
\centering
\caption{Summary of descriptive results of user performance.}
\begin{tabular}{lrrr}
\hline
Variable & Task Completion Time (s) & Normalized Number of Mistakes & Resistance Change (\%) \\
\hline
VR & 26.60 (14.10) [2.50] & 0.088 (0.16) [0.03] & 605.11 (546.43) [99.76] \\
\hspace*{0.2cm} Circle & 26.29 (15.85) [3.96] & 0.088 (0.10) [0.02] & 486.99 (354.49) [91.53] \\
\hspace*{0.2cm} Diamond & 26.99 (12.66) [3.16] & 0.089 (0.21) [0.05] & 723.23 (680.40) [175.68] \\
VR KinArm & 24.10 (8.59) [1.52] & 0.134 (0.19) [0.03] & 579.69 (525.03) [95.85] \\
\hspace*{0.2cm} Circle & 24.67 (8.84) [2.21] & 0.088 (0.17) [0.04] & 562.62 (529.04) [136.60] \\
\hspace*{0.2cm} Diamond & 23.45 (8.56) [2.14] & 0.180 (0.20) [0.05] & 596.76 (538.97) [139.16] \\
\hline
\end{tabular}
\\
\textit{All entities are in the format: mean value (standard deviation) [standard error].}
\label{tab:descriptiveResults}
\end{table*}

\begin{table}[!t]
\centering
\caption{Summary of statistical results ($p<.05$).}
\resizebox{\columnwidth}{!}{
\begin{tabular}{lllrrcr}
\hline
Variable  & DFn & DFd & F & p & Sig. & $\eta_{p}^{2}$  \\
\hline
Task Completion Time & & & & & & \\
\hspace*{0.2cm} Condition & 1 & 15 & 0.921 & 0.352 &  & 0.012 \\
\hspace*{0.2cm} Task & 1 & 15 & 0.039 & 0.845 &  & 0.0001 \\
\hspace*{0.2cm} Condition * Task & 1 & 15 & 0.484 & 0.496 &  & 0.001 \\
Norm. Number of Mistakes & & & & & & \\
\hspace*{0.2cm} Condition & 1 & 15 & 4.290 & 0.055 &  & 0.017 \\
\hspace*{0.2cm} Task & 1 & 15 & 4.983 & 0.041 & * & 0.017 \\
\hspace*{0.2cm} Condition * Task & 1 & 15 & 1.098 & 0.311 &  & 0.016 \\
Resistance Change & & & & & & \\
\hspace*{0.2cm} Condition & 1 & 14 & 0.037 & 0.849 &  & 0.0005 \\
\hspace*{0.2cm} Task & 1 & 14 & 5.319 & 0.036 & * & 0.016 \\
\hspace*{0.2cm} Condition * Task & 1 & 14 & 4.287 & 0.057 &  & 0.009 \\
\hline
\end{tabular}
\label{tab:anovaresults}
}
\end{table}

\subsubsection{Normalized Number of Mistakes}
We found no significant difference of the \textit{Condition} factor. 
However, there was a significant difference of the \textit{Task} factor between \textit{Circle} and \textit{Diamond} (\textit{F}(1, 15) = 4.983, \textbf{\textit{p} = 0.041}, $\eta_{p}^{2}$ = 0.017; small effect). 
We further analyzed the data with pairwise t-test, and found the significant effect (\textit{t} = -1.92, \textit{df} = 15, \textbf{\textit{p} = 0.037}) between the \textit{Circle} (\textit{M} = 0.088, \textit{SD} = 0.04) and \textit{Diamond} (\textit{M} = 0.180, \textit{SD} = 0.05) in the \textit{VR KinArm Condition} -- 
the raw number of mistakes before normalization: \textit{Circle} (\textit{M} = 0.812, \textit{SD} = 1.51) and \textit{Diamond} (\textit{M} = 1.688, \textit{SD} = 1.49). 
This could indicate that the participants made more mistakes with the \textit{Diamond Task} in the \textit{VR KinArm Condition}. It is also noteworthy that the task completion time for this task was faster than others. Thus, it could show that the participants 
tried to perform this task quickly but made more mistakes. 
There was no significant effect on the interaction effect of both \textit{Condition} and \textit{Task} factors.

\subsubsection{Resistance Change}

There was no significant difference on the \textit{Condition} factor. In the \textit{Task} factor, however, we found a significant effect (\textit{F}(1, 14) = 5.319, \textbf{\textit{p} = 0.036}, $\eta_{p}^{2}$ = 0.016; small effect).
The results of pairwise t-test show a significant difference (\textit{t} = -2.35, \textit{df} = 14, \textbf{\textit{p} = 0.034}) between the \textit{Circle} (\textit{M} = 486.99, \textit{SD} = 354.49) and \textit{Diamond} (\textit{M} = 723.23, \textit{SD} = 680.40) \textit{Tasks} in the \textit{VR} condition. 
This indicates that during the \textit{Diamond} task, the participants bent their elbow more, leading to a higher resistance change, when compared to the \textit{Circle} task in \textit{VR} condition. 
In the \textit{VR KinArm} condition, however, there was a small difference between the \textit{Circle} (\textit{M} = 562.62, \textit{SD} = 529.04) and \textit{Diamond} (\textit{M} = 596.76, \textit{SD} = 538.97). 
We did not observe a significant effect on the interaction effect between the factors.

\begin{figure*}[t]
    \centering    
    \includegraphics[width=\textwidth]{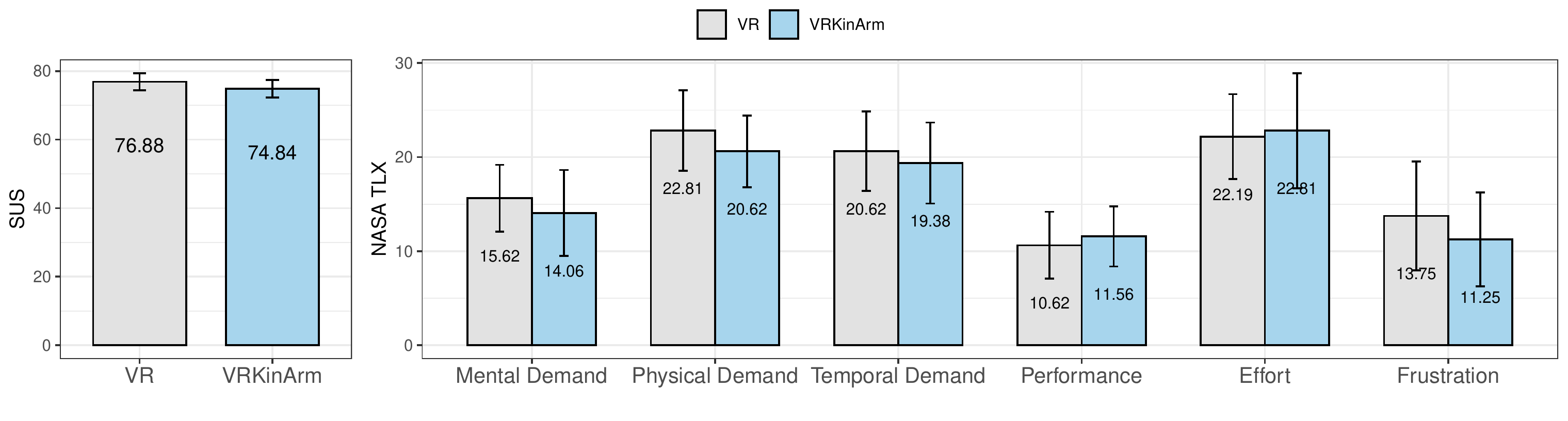}
    \caption{Questionnaire results: (left) system usability scale (SUS) and (right) NASA task load index (TLX).}
 \label{fig:questionnaireResults}
\end{figure*}

\subsection{Questionnaire Results (RQ2)}

We describe the subjective results from the questionnaires in the following sections.

\subsubsection{Usability}

The results of system usability using SUS questionnaire from all participants show an average score of (\textit{M} = 76.88, \textit{SD} = 10.10) for \textit{VR} and (\textit{M} = 74.84, \textit{SD} = 10.18) for \textit{VR KinArm Condition} (see \autoref{fig:questionnaireResults}). 
Descriptive results show that there was a small difference between both \textit{Conditions}. 
However, we did not find any significant differences on the main effect and their interaction effect.
The SUS scores for both \textit{Conditions} higher than 68 are above average, which could indicate their potential benefits in terms of usability \citep{bangor2009determining}.

\subsubsection{Task Load}

The subjective task load among \textit{Conditions} was assessed using unweighted (raw) NASA-TLX questionnaire.  
Descriptive results for overall scores show an average of (\textit{M} = 17.60, \textit{SD} = 17.68) for \textit{VR} and (\textit{M} = 16.61, \textit{SD} = 18.44) for \textit{VR KinArm Condition}. 
Moreover, descriptive results show that the average scores of frustration, mental, physical, and temporal demands in the \textit{VR} are averagely higher, while the performance and effort are averagely lower than the \textit{VR KinArm Condition} (see \autoref{fig:questionnaireResults}). 
However, the scores are slightly different. 
We found no significant differences between two \textit{Conditions} for all the task load items. 
The results could show that the participants had similar task loads, and overall scores below 20 are still considered as low for two \textit{Conditions}.  

\subsubsection{Presence}

\begin{figure}[!t]
    \centering    
    \includegraphics[width=\columnwidth]{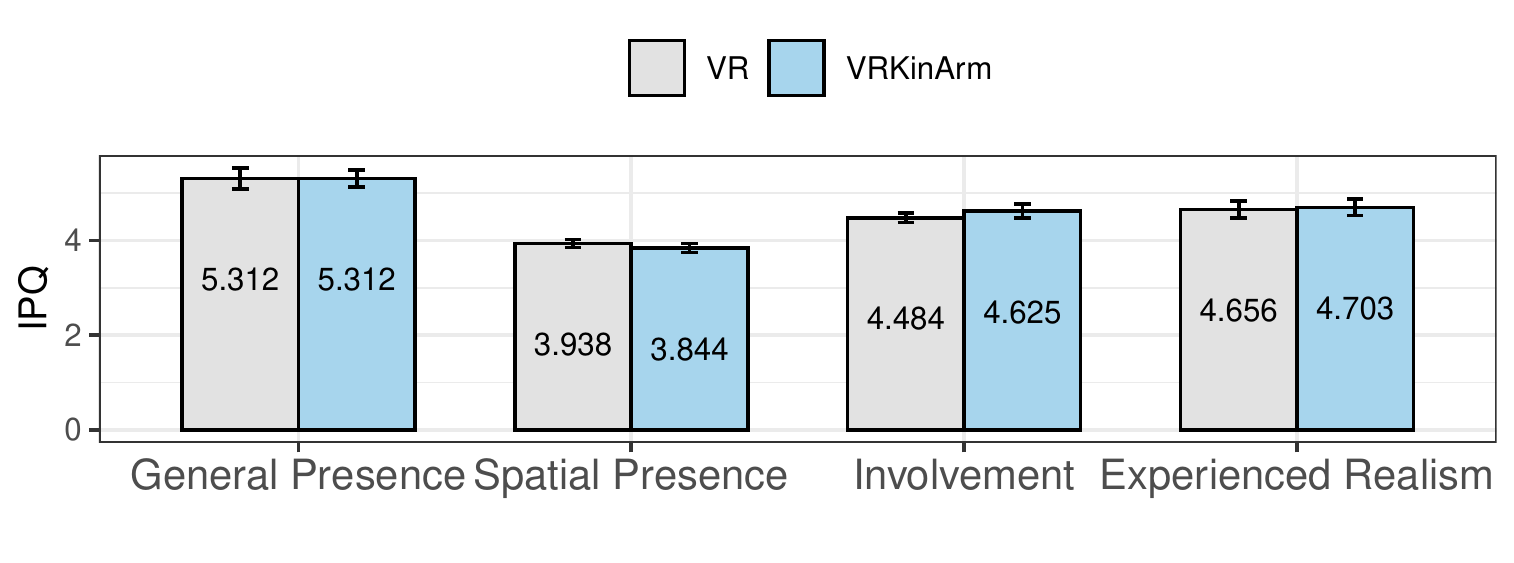}
    \caption{Results of \textit{Igroup} presence questionnaire (IPQ).}
 \label{fig:ipqResults}
\end{figure}

We assessed the sense of presence in the immersive environment using an IPQ questionnaire (see \autoref{fig:ipqResults}). 
There were no significant differences among two \textit{Conditions}. 
It is noteworthy that the general presence has identical average scores for \textit{VR} (\textit{M} = 5.31, \textit{SD} = 0.87) and \textit{VR KinArm} (\textit{M} = 5.31, \textit{SD} = 0.70) condition. 
The spatial presence of the \textit{VR Condition} (\textit{M} = 3.94, \textit{SD} = 0.33) is averagely higher than \textit{VR KinArm} (\textit{M} = 3.84, \textit{SD} = 0.38), while the involvement (\textit{M} = 4.48, \textit{SD} = 0.41) and experienced realism (\textit{M} = 4.66, \textit{SD} = 0.70) in the \textit{VR Condition} are on the average lower than the \textit{VR KinArm}: involvement (\textit{M} = 4.63, \textit{SD} = 0.57) and experienced realism (\textit{M} = 4.70, \textit{SD} = 0.67). 
The descriptive results, however, show slight differences between the conditions.   

\subsection{Qualitative Participant Feedback (RQ3)}

In the following, we summarize the participants' feedback during the semi-structured interview.

All participants were positive regarding using VR and end-point robots for upper extremity rehabilitation. Six participants clearly stated that it could be used to engage and motivate patients to exercise, especially for upper extremity therapy, through the immersive experience. 
Moreover, it benefits patients with limited mobility because the virtual content and surrounding environments can be customized. 
Two participants with no previous VR experience mentioned that the VR headset was quite heavy for them, and they stated that it could also be an issue for the patients. Thus, using lightweight devices would be more comfortable. 
They also stated that the computer-generated environment and dots were unclear and flickering.  
Nonetheless, other participants described that it was a good experience using the system. 

Five participants expressed that the \textit{Tasks} were not difficult. They appreciated that the aim was also for elbow assessment, and it was in an early stage of system development.
However, investigating real scenarios for rehabilitation from the clinic would be interesting for further development.
One participant suggested enlarging the white dot, which indicates the position of the VR tracker on the screen. 

Regarding \textit{Conditions}, four participants stated that using the VR tracker in the \textit{VR Condition} was easy and portable to hold. Moreover, it is advantageous for \textit{home therapy} with its portability and cost-effectiveness.  
Five participants expressed their thoughts about using \textit{VR} and KinArm robot (\textit{VR KinArm}). 
One 
advantage is that it is steady and provides haptic feedback. 
Combining the KinArm robot with VR requires calibration with a VR setup to fit in with the virtual KinArm in the virtual environment. Four participants stated there were tracking issues, especially during the training phase, which required some time to match them with the virtual environment.   
Two participants suggested using collaborative (multi-user) VR for home therapy because it could provide benefits ranging from increased enjoyment to telerehabilitation. Another participant recommended changing the 
rubber band used on the wearable sleeve sensor.

\section{Discussion}

In this section, we summarize the main findings and discuss the implications of results described in the previous section.

Our proposed system for upper extremity rehabilitation using VR and robotics was evaluated in a pilot study. Objective measures based on user performance, subjective measures based on questionnaires, and qualitative feedback were measured and reported. 
For user performance, there was no significant difference among both factors (\textit{Condition} and \textit{Task}) and their interaction effect. 
The results could indicate that the setup for \textit{VR} (using VR tracker) and \textit{VR KinArm} (using KinArm end-point handle) are comparable, thus interchangeable in terms of task completion time. 
This can mean that the VR representation of the KinArm can produce the same performance results, which can be beneficial to patients since the VR system is more accessible and home-based than a VR KinArm system. 
However, based on the slight trends in descriptive results, the \textit{VR KinArm Condition} is advantageous. 
Some participants stated that using the \textit{VR KinArm} provides a steady movement and allows them to perform the tasks faster. 
The results of normalized number of mistakes, however, showed that there was a significant difference between the \textit{Tasks} in the \textit{VR KinArm Condition}. Participants made more mistakes performing \textit{Diamond Task} within this \textit{VR KinArm Condition}. It is noteworthy that this task was performed faster than others. It could indicate that \textit{VR KinArm Condition} provided a steady movement, and the participants attempted to perform it quickly.

No significant difference in the main effect was observed between the \textit{VR} and \textit{VR KinArm} conditions for the resistance change.  
This indicates that two \textit{Conditions} of the study produced comparable outcomes for resistance change measured at the elbow using a wearable sensor.
However, there was a significant difference in the \textit{Task} factor, which shows that the participants applied more resistance changes on the \textit{Diamond} task in the \textit{VR Condition}. This resistance change values for two \textit{Tasks} underscores how the wearable sensor is sensitive enough to successfully distinguish study \textit{Tasks} from each other. The \textit{Circle Task} has a curved shape which results in lower resistance change values being recorded from participants compared to the \textit{Diamond Task}, which has sharper trajectories and 
leads to more bending of elbow causing a higher change in resistance. 
This difference in resistance change may be due to the shape of the task and the type of movement required to complete it. 
Descriptive results also show that the resistance change measured during the \textit{Diamond} task is higher in both \textit{Conditions}. 
Nonetheless, no significant difference in the interaction effect was found. 
The use of wearable sensors with \textit{VR} setup enables the accurate measurement of human movements outside the line of sight of VR trackers. These sensors are extremely sensitive, breathable, and comfortable to wear, making them non-invasive and convenient to wear for patients during therapy. Using VR therapy with wearable sensors can potentially provide accurate data about a patient's progress and compliance. Challenges with respect to calibration, scalability of sensor manufacturing and data protocols, processing, and protection are being addressed by researchers.

We assessed the subjective measures of the system usability, task load, and sense of presence with the standardized questionnaires.  
The results of the SUS questionnaire reveal both benefits of two \textit{Conditions} in terms of usability. 
The task load between the \textit{Conditions} was assessed using NASA-TLX questionnaire. The results of overall scores indicate that the task load for both \textit{Conditions} is relatively low. Moreover, we observed no significant effect among the items.
For descriptive results, the physical demand was rated higher than other items in the \textit{VR Condition}, while the effort was rated the highest in the \textit{VR KinArm Condition}. 
The results show the advantages of both \textit{Conditions} regarding the task load. 

We used the IPQ questionnaire to assess the sense of presence in the immersive environment.
It is also noteworthy that nine among 16 participants had no previous VR experience.
The results show that experienced realism was rated as the highest among other scales of IPQ for both conditions. It could indicate that the system provides a good experience and realism of the environment. In contrast, the spatial presence was rated as the lowest. One reason might be that the physical KinArm robot was placed in front of the participants, reducing the degrees of freedom. However, they reported that they still had a sense of presence while experiencing the immersive environment.    
The general presence and involvement, which measures the awareness and attention in the virtual environment, were rated as a good, suitable, and reasonable environment for VR experience.

\paragraph{Limitations and Future Directions}

This initial study included a limited number of non-clinical participants. Our goal is to extend the results to rehabilitating upper limbs in a clinical population, such as stroke patients or those with Parkinson's disease. 
The task designed for the study was a VR drawing game focused on the upper extremity condition, and it was also aimed to provide a valid assessment of elbow joint movement. 
In the future, exploring various therapeutic interventions in VR will be beneficial, including those that involve joint movements in the lower limb for rehabilitation purposes \citep{doshi2019carbon}. 
We integrated a KinArm end-point robot with the immersive VR setup; however, there were several challenges to providing a better tracking space for the VR experience. One solution is to remove the upper portion (display) of the KinArm robot and use only the robot handles to perform the task. Hence, investigating portable robotics to integrate with VR would be interesting for future work.     
Moreover, comparing different interaction techniques and input modalities, for instance, hands-free vs VR controller interactions would be interesting as well \citep{juan2022immersive}. 
Another future research is collaborative, multi-user VR, which allows seamless collaboration between patients and therapists \citep{thielbar2020utilizing, chheang2022towards, chheang2021collaborative}. 


\section{Conclusion}

We have presented a framework for upper extremity rehabilitation using immersive VR and robotics. We conducted a pilot user study and assessed user performance and movement using a wearable sleeve sensor made of knitted nanotube fabric. Results of subjective evaluations obtained from self-administered questionnaires, including usability, task load, and presence, were reported.
The proposed framework demonstrates the potential advantages of an immersive, multi-sensory approach and provides future avenues for research in developing more cost-effective and personalized upper extremity rehabilitation solutions.



\bibliographystyle{ACM-Reference-Format}
\balance
\bibliography{Main.bib}


\end{document}